\documentclass[aps,pra,superscriptaddress,twocolumn,nofootinbib]{revtex4-1}

\usepackage{graphicx,graphics,epsfig}   
\usepackage{dcolumn}    
\usepackage{bm}         
\usepackage{amsmath}    
\usepackage{verbatim}   
\usepackage{color}      
\usepackage{subfigure}  
\usepackage{amsmath,amsfonts,amssymb,amsthm,graphics,graphics,color,bbm}
\usepackage[rgb]{xcolor}
\usepackage{enumerate, tikz}

\usetikzlibrary{calc,decorations.pathreplacing}
\usetikzlibrary{arrows,shapes}

\newcommand{\tr}[1]{\mathrm{tr} \left\{ #1 \right\}}
\newcommand{\ket}[1]{|#1\rangle}
\newcommand{\bra}[1]{\langle#1|}
\newcommand{\braket}[2]{\langle#1|#2\rangle}

\newcommand{\expect}[1]{\langle#1\rangle}

\newcommand{\Q}{\mathcal{Q}}
\renewcommand{\P}{\mathcal{P}}
\newcommand{\AQ}{\widetilde{\mathcal{Q}}}

\newcommand{\NS}{\mathcal{NS}}

\newcommand{\vecP}{\vec{p}}
\newcommand{\tp}{^\text{\tiny $T_i$}}

\newcommand{\be}{\begin{eqnarray}}
\newcommand{\ee}{\end{eqnarray}}

\definecolor{nred}{rgb}{0.9,0.1,0.1}
\definecolor{nblack}{rgb}{0,0,0}
\definecolor{nblue}{rgb}{0.2,0.2,0.8}
\definecolor{ngreen}{rgb}{0.2,0.6,0.2}

\usepackage{hyperref}

\renewcommand{\L}{\mathcal{L}}

\begin{document}

\title{Almost quantum correlations and their refinements in a tripartite Bell scenario}

\author{James Vallins}
\affiliation{H. H. Wills Physics Laboratory, University of Bristol$\text{,}$ Tyndall Avenue, Bristol, BS8 1TL, United Kingdom}
\author{Ana Bel\'en Sainz}
\affiliation{H. H. Wills Physics Laboratory, University of Bristol$\text{,}$ Tyndall Avenue, Bristol, BS8 1TL, United Kingdom}
\affiliation{Perimeter Institute for Theoretical Physics, 31 Caroline St. N, Waterloo, Ontario, Canada N2L 2Y5}
\author{Yeong-Cherng Liang}
\affiliation{Department of Physics, National Cheng Kung University, Tainan 701, Taiwan}

\begin{abstract}
We study the set of almost quantum correlations and their refinements in the simplest tripartite Bell scenario where each party is allowed to perform two dichotomic measurements. In contrast to its bipartite counterpart, we find that there already exist {\em facet} Bell inequalities that witness almost quantum correlations beyond quantum theory in this simplest tripartite Bell scenario. Furthermore, we study the relation between the almost quantum set and the hierarchy of supersets to the quantum set due to Navascu\'es-Pironio-Ac\'in (NPA) [Phys. Rev. Lett. {\bf 98}, 010401 (2007)]. While the former lies between the first and the third level of the NPA hierarchy, we find that its second level does not contain and is not contained within the almost quantum set. Finally, we investigate the hierarchy of refinements to the almost quantum set due to Moroder {\em et al.} [Phys. Rev. Lett. {\bf 111}, 030501 (2013)], which converges to the set of quantum correlations producible by quantum states having positive partial transposition. This allows us to {consider (approximations of)} the ``biseparable" subsets of the almost-quantum set as well as of the quantum set of correlations,  and thereby gain further insights into the subtle similarities and differences between the two sets. In addition, they allow us to identify candidate Bell-like inequalities that can serve as device-independent witnesses for genuine tripartite entanglement.
\end{abstract}
\date{\today}

\maketitle

One of the most striking discoveries of the last century is that Nature is incompatible with classical physics. Powerful correlations which cannot be explained in the classical world do arise in Nature, and are accurately predicted by quantum mechanics~\cite{clauser, aspect, monroe, delft, nist, zeil}. This phenomenon, known as Bell-nonlocality~\cite{RMP} (or more often simply abbreviated as nonlocality), was put on firm theoretical grounds by Bell~\cite{bell} in 1964, almost three decades after Einstein, Podolsky and Rosen~\cite{epr} first noticed the phenomenon of quantum entanglement~\cite{Ent}. The discovery of these nonclassical correlations not only imposed a change in the way we may understand Nature, but also leads to the discovery of powerful resources for information tasks. For instance, quantum correlations with no classical analog have been proven essential for device-independent quantum key distributions \cite{qkd1,qkd2} and device-independent random number generations \cite{ran1,ran2}. 

Quantum mechanics, however, does not exploit the phenomenon of nonlocality to its maximum, i.e, even stronger correlations conforming to the assumption of relativistic causality~\cite{Tsi80,PR94} (or more often referred to as the no signalling condition~\cite{Barrett05}) are in principle allowed. The natural question then is whether there is a way to understand from basic principles why such correlations beyond quantum mechanics should not arise in Nature. In the past decades several physical and information-theoretical principles have been proposed \cite{ntcc,ntcc2,IC,nanlc,ML,LO, EP, MNC}. {While they allow for a better understanding of quantum correlations, most of them are provably satisfied by a superset of it, dubbed {\it almost quantum} \cite{aq}, which includes postquantum ones.} The only exception\footnote{{There is also the Exclusivity principle \cite{EP}, which imposes constraints on physical theories that predict correlations. Therefore, how to apply it to the almost quantum ones is still a topic of discussion.}} to this is the principle of Information Causality (IC) \cite{IC}, where there is so far only numerical evidence hinting that the almost quantum correlations also satisfy it \cite{aq}. 

In this work, we shall indeed focus our attention on this intriguing set of postquantum correlations~\cite{aq, npa,Moroder:PRL:2013}. Almost quantum correlations have been widely studied in the bipartite Bell scenarios. For the case where each experimenter chooses between two dichotomic measurements, almost quantum correlations violate the renowned Clauser-Horne-Shimony-Holt~\cite{CHSH} (CHSH) Bell~\cite{bell} inequality to the same extent as the quantum ones, so in order to tell these two sets apart in this simplest setting, {\em non-facet}~\cite{RMP} Bell inequalities such as those discussed in~\cite{Liang:PRA:2011} are required\footnote{{Given an almost quantum correlation with no quantum realisation, there are other theoretical ways to signal its postquantumness. For example, one could test membership to the NPA sets $\Q_k$ (see Appendix \ref{App:NPA}) and find a $k$ that does not contain it (see also page 95 of Ref.~\cite{Wolfe:thesis}). However, evaluating a Bell inequality is the traditional way that experiments are carried out, and so we focus on that method in this work.}}. When considering three measurements instead, both the facet Bell inequality I$_{3322}$ and I$_{3333}$ from Ref.~\cite{Collins:JPA:2004} display a gap between quantum and almost quantum violations (see, respectively, \cite{npa} and \cite{Schwarz:NJP:2015}); however, no physical principle (to bound correlations) has been distilled from these inequalities so far.  

Here, one of the next simplest Bell scenarios is considered, namely, that consisting of three parties performing two dichotomic measurements each. The set of classical correlations in this scenario is fully characterised by 46 classes of facet Bell inequalities \cite{sliwa}. Among these are the Mermin \cite{mermin} and the tripartite {\it guess your neighbour's input} (GYNI) \cite{gyni} inequalities, which display curious properties. On the one hand, the maximum value of Mermin's inequality for no-signalling (NS) correlations coincides with that for quantum (and hence almost quantum) correlations, while classical ones achieve a lower value. On the other hand, for the GYNI inequalities,  the classical maximum coincides with that for the almost quantum (and hence the quantum) correlations \cite{LO}, thereby giving them the interpretation as games with no quantum advantage, while the NS ones achieve a larger value. Therefore, the range of possibilities that this tripartite scenario displays is much richer than its bipartite counterpart. 

In this work, first we ask in Sec.~\ref{Sec:Local} the question of whether we can tell apart the sets of quantum and almost quantum correlations by only looking at their violation of facet Bell inequalities. We will see that this rich yet simple tripartite Bell scenario allows us to do so. Then, in Sec.~\ref{Sec:NPA}, we move on to study the relations between the set of almost quantum correlations and those sets of postquantum correlations defined by Navascu\'es, Pironio and Ac\'in (NPA)\footnote{For bipartite scenarios, the set of almost quantum correlations was already introduced in the original NPA papers~\cite{npa} as an `intermediate set' between the first and second levels of the hierarchy. And indeed, one can always understand almost quantum correlations as an `intermediate NPA set'. In this work, however, when studying the NPA hierarchy we will focus on canonical generating sets (see Appendix \ref{App:NPA}).} \cite{npa}. In particular, we show that in the tripartite, and therefore in a {general $n$-partite scenario (where $n\ge 3$), neither} the $(n-1)$-th level of the NPA hierarchy is contained within the almost quantum set, nor otherwise. Lastly, we discuss in Sec.~\ref{Sec:Refine} another hierarchy of supersets to the quantum set due to Moroder~{\em et al.}~\cite{Moroder:PRL:2013}. Since the lowest level of this hierarchy corresponds precisely to the almost quantum set~\cite{aq,miguel}, higher levels in this hierarchy can be seen as refinements to the almost quantum set. By focusing ourselves on further refinements that can {be conceived as (approximations) to the ``biseparable" subsets} of the almost quantum set, we gain further insight into some subtle features of the almost quantum set. We conclude in Sec.~\ref{Sec:Conclusion} with some possibilities for future work.

\section{The simplest tripartite Bell scenario}

\begin{figure}
\begin{center}
\begin{tikzpicture}[scale=0.9]

\shade[draw, thick, ,rounded corners, inner color=white,outer color=gray!50!white] (-3.25,0.7) rectangle (-1.75,2) ;
\shade[draw, thick, ,rounded corners, inner color=white,outer color=gray!50!white] (-0.75,0.7) rectangle (0.75,2) ;
\shade[draw, thick, ,rounded corners, inner color=white,outer color=gray!50!white] (1.75,0.7) rectangle (3.25,2) ;

\draw[thick, ->] (0,2.7) -- (0,2);
\draw[thick, ->] (-2.5,2.7) -- (-2.5,2);
\draw[thick, ->] (2.5,2.7) -- (2.5,2);

\draw[thick, ->] (0,0.7) -- (0,0);
\draw[thick, ->] (-2.5,0.7) -- (-2.5,0);
\draw[thick, ->] (2.5,0.7) -- (2.5,0);

\node at (-2.5,3.75) {Alice};
\node at (2.5,3.75) {Charlie};
\node at (0,3.75) {Bob};
\node at (-2.5,3) {$x$};
\node at (-2.5,-0.3) {$a$};
\node at (0,3) {$y$};
\node at (0,-0.3) {$b$};
\node at (2.5,3) {$z$};
\node at (2.5,-0.3) {$c$};
\draw[thick, decoration={brace},decorate] (3.3,-0.7) -- (-3.3,-0.7);
\node at (0, -1.5) {$p(abc|xyz)$};
\end{tikzpicture}
\end{center}
\caption{A tripartite Bell scenario. Three parties (Alice, Bob and Charlie) can operate locally each on their share of a system, depicted as a black box. They choose measurements labelled by $x$, $y$ and $z$ and register the obtained outcomes $a$, $b$ and $c$, respectively. The correlations between measurement outcomes observed are captured by the vector of conditional probability distributions $\vecP=\{p(abc|xyz)\}$. }\label{Bellsce}
\end{figure}
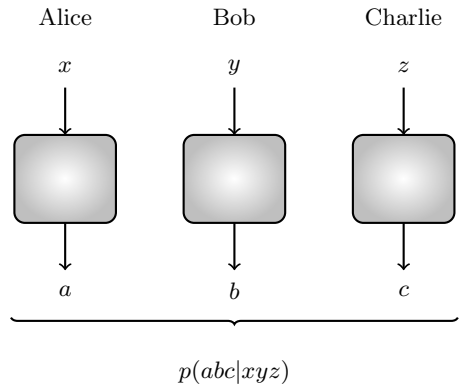

A tripartite Bell scenario (see Fig.~\ref{Bellsce}) consists of three parties (Alice, Bob and Charlie) each of which can act locally on their share of a system. The parties are assumed to be in spatially separated locations, and no information can travel from one party to another during the experiment. Otherwise, no assumptions are made on the systems the parties possess or the specific way the measurements are implemented. That is, the only information the parties have access to is the classical labels of their measurement choices $x$, $y$ and $z$ and outcomes $a$, $b$ and $c$, respectively for Alice, Bob and Charlie. The correlations between measurement outcomes are then succinctly summarized as a collection of conditional probabilities $\vecP=\{p(abc|xyz)\}$. {Hereafter,} we focus on the simplest case where the parties can perform two dichotomic measurements, i.e., $x,y,z \in \{0,1\}$ and $a,b,c \in \{0,1\}$.

Classical (i.e., Bell-local~\cite{RMP}) conditional probability distributions are those where the correlations among the parties can only be caused by a common classical agent $\lambda$, referred to as `shared randomness' or `local hidden variables'. These correlations can be written as $p(abc | xyz) = \int {d\lambda}\,q(\lambda) p_\mathrm{A}(a|x,\lambda) \, p_\mathrm{B}(b|y,\lambda) \, p_\mathrm{C}(c|z,\lambda)$, where $q(\lambda)$ is a normalised distribution over the shared randomness and $p_\mathrm{A}(a|x,\lambda)$ is a well-defined conditional probability distribution for Alice for each $\lambda$ (similarly for Bob and Charlie). Hereafter, we denote the set of Bell-local correlations by $\L$.

The set of quantum correlations $\Q$ are those that arise by the parties performing measurements on a shared quantum state (possibly entangled). A given correlation $\vecP$ belongs to the quantum set if there exists a normalised quantum state $\rho$ and two projective measurements per party, $\{\Pi_{0|x}, \Pi_{1|x} = \mathbbm{1} - \Pi_{0|x} \}_{x=0,1}$ (similarly for Bob and Charlie) such that the correlations arise via Born's rule: $p(abc|xyz) = \tr{\Pi_{a|x} \, \Pi_{b|y} \, \Pi_{c|z} \, \rho}$. One important property that these projectors should satisfy is that those corresponding to different parties commute, i.e., $[\Pi_{a|x},\Pi_{b|y}] = 0$ $\forall \, a,b,x,y$ (similarly for the other two combinations of parties). Note that there are no restrictions on the dimension of the Hilbert space.

The set of almost quantum correlations~\cite{aq}, denoted by $\tilde{\Q}$, can be understood as a relaxation of the quantum set in the following sense. A given $p(abc|xyz)$ belongs to the almost quantum set if there exists a normalised quantum state $\rho$ and two projective measurements per party, $\{\Pi_{0|x}, \Pi_{1|x} = \mathbbm{1} - \Pi_{0|x} \}_{x=0,1}$ (similarly for Bob and Charlie) such that the correlations arise via Born's rule: $p(abc|xyz) = \tr{\Pi_{a|x} \, \Pi_{b|y} \, \Pi_{c|z} \, \rho}$. However, here we do not demand that the projectors corresponding to different parties commute, {instead we impose the following: $\Pi_{a|x} \, \Pi_{b|y} \, \Pi_{c|z} \, \rho \, (\Pi_{a|x} \, \Pi_{b|y} \, \Pi_{c|z})^\dagger = \Pi \, \rho \, \Pi^\dagger$, where $\Pi$ is any permutation of $\Pi_{a|x} \, \Pi_{b|y} \, \Pi_{c|z}$ (a nontrivial permutation, for instance, is given by $\Pi = \Pi_{a|x} \, \Pi_{c|z} \Pi_{b|y} \, $, which swaps the projectors $\Pi_{b|y}$ and $\Pi_{c|z}$).} As this last requirement on the operators does not ensure that projectors associated with different parties, such as $\Pi_{a|x}$ and $\Pi_{b|y}$, commute, it thus follows that the set of quantum correlations is a subset to the almost quantum set.

Finally, the set of no-signalling correlations $\NS$ are defined as those $\vecP$ such that all their marginals are well-defined, i.e., they do not depend on the measurement choices of the parties that are traced out, and therefore do not allow for faster-than-light communication. More precisely, {$\sum_{c} p(abc|xyz)$ should be independent of $z$, $\sum_{bc} p(abc|xyz)$} should be independent of $y,z$, and analogous conditions should hold for other marginal distributions.

\section{The almost quantum set and the tripartite local polytope}
\label{Sec:Local}

In this simplest Bell scenario that we are considering, the set of local correlations is fully characterised by 53\,856 tight (i.e., {\em facet}) Bell inequalities~\cite{Pitowsky:PRA:2001}. In other words, a correlation $\vecP$ is local if and only if it satisfies all these distinct Bell inequalities. These inequalities can be classified into 46 inequivalent\footnote{Following Ref.~\cite{Collins:JPA:2004}, we say that two Bell inequalities are inequivalent if they cannot be obtained from one another via the relabelling of measurement settings, outcomes and parties.} classes~\cite{sliwa}. Hereafter, we will focus on the representative of each of these classes as provided by Sl\'iwa in Ref.~\cite{sliwa}. Specifically, we are interested in the maximum violation that almost quantum correlations yield for these representative inequalities and whether any of these {\em cannot} be achieved by $\vecP\in\Q$. 

Computing the maximum violation of a Bell inequality for correlations restricted to the set of almost quantum correlations turns out to be a semidefinite program (SDP) (see Appendix~\ref{App:AQ}), which is a kind of convex optimization problem that can be efficiently solved on a computer.\footnote{This can be achieved with the help of a modeling software such as CVX~\cite{cvx} or YALMIP~\cite{yalmip}, in conjunction with an SDP solver, such as sedumi~\cite{sedumi} or SDPT3~\cite{sdpt3}.}  Likewise, computing an upper bound on the maximal quantum violation can also be achieved by solving, for instance, the hierarchy of SDPs introduced by NPA~\cite{npa} {(see Appendix~\ref{App:NPA}) or Moroder {\em et al.}~\cite{Moroder:PRL:2013} (see Appendix~\ref{App:AQ}).} To ensure that the upper bound obtained is tight (i.e., indeed corresponds to the maximal violation allowed by quantum theory), we also perform iterative optimisation by considering local projective measurements on three-qubit states, which is known~\cite{lluis} to be sufficient in this Bell scenario. For all of the 45 nontrivial (representative) facet Bell inequalities of this scenario, the best lower bounds that we have obtained using standard iterative optimisation techniques (see, e.g., Ref.~\cite{Opt}), match with the corresponding upper bound obtained by solving the aforementioned SDP
(up to the numerical precision of the solver\footnote{This is of the order of $10^{-8}$.}).

Our results are presented in Table~\ref{Bounds322}, following the labelling of Ref.~\cite{sliwa}. Out of the 45 nontrivial inequalities, 43 display the same phenomenon as in the bipartite CHSH scenario (up to {the numerical precision of the solver}): the maximum violation {attainable by the almost quantum correlations (column four, Table~\ref{Bounds322}) can also be achieved by the quantum ones (column three, Table~\ref{Bounds322})}. However, two inequalities (n$^o$ 23 and 41) display a gap, that is, they demonstrate that quantum correlations cannot be as nonlocal as the almost-quantum ones. This is the first time that such a behaviour is observed for scenarios with two dichotomic measurements per party using facet Bell inequalities. The particular form of these two inequalities is: 

\begin{widetext}
\begin{align}
&\text{Ineq. 23:} \nonumber\\
&E(x=0)+E(x=1)+E(y=0)-E(x=0,y=0)-E(x=1,y=0)+E(y=1)-E(x=0,y=1)\\ 
&-E(x=1,y=1)+E(x=0,z=0)-E(x=1,z=0)-E(000)+E(100)-E(010)+E(110) \nonumber\\
& +E(y=0,z=1)-E(001)-E(101)-E(y=1,z=1)+E(011)+E(111) \stackrel{\L}{\leq} 4\,,\nonumber
\end{align}
\begin{align}
&\text{Ineq. 41:} \nonumber\\
&E(x=0)+E(y=0)+E(x=0,y=0)+E(z=0)+E(x=1,z=0)-3E(000)-E(100)+E(y=1,z=0)\\
&-E(010)-2E(110)+E(x=0,z=1)-E(x=1,z=1)+E(y=0,z=1)-4E(001)+E(101) \nonumber \\
&-E(y=1,z=1)+E(011)+2E(111) \stackrel{\L}{\leq} 7 \,, \nonumber
\end{align}
\end{widetext}
where $\L$ signifies that the inequality holds for all correlations $\vecP\in\L$, while $E$ are single, two and three-body correlators defined as follows: $E(x) = p_A(0|x)-p_A(1|x)$ (similarly for Bob and Charlie), $E(xy) = \sum_{ab} (-1)^{a+b} p_{AB}(ab|xy)$ (similarly for the other pairs of parties) and $E(xyz) = \sum_{abc} (-1)^{a+b+c} p(abc|xyz)$. Here $p_{AB}(ab|xy) = \sum_c p(abc|xyz)$ and $p_A(a|x) = \sum_{b,c} p(abc|xyz)$, which are well-defined marginals for no-signalling correlations. 

\begin{table*}
\begin{tabular} {|c|c|c|c|c|c||c|c|c|c||c|c|c|c|}
\hline
Inequality & $\L$ & $\Q$  & $\AQ$ & $\Q_2$ & $\NS$ & $\AQ_{1}^\text{\tiny $T_A$}$ & $\AQ_{1}^\text{\tiny $T_B$}$ & $\AQ_{1}^\text{\tiny $T_C$}$ & $\AQ_{1}^\text{\tiny $T_{all}$}$ & $\AQ_{6}^\text{\tiny $T_A$}$ & $\AQ_{6}^\text{\tiny $T_B$}$ & $\AQ_{6}^\text{\tiny $T_C$}$ & $\AQ_{6}^\text{\tiny $T_{all}$}$ \\
\hline
1 & 1 & 1 & 1 & 1 & 1 & 1 & 1 & 1 & 1 & 1 & 1 & 1 & 1 \\
2 & 2 & 4 & 4 & 4 & 4  &   2.8284 &   2.8284 &   2.8284 &   2.0000 & 2.8284 & 2.8284 &    2.8284 &    2.0000\\
3 & 2 & 2.8284 & 2.8284 & 2.8284 & 4 &   2.0000 &   2.8284 &   2.8284 &   2.0000 & 2.0000 &    2.8284 &    2.8284 &    2.0000 \\
4 & 2 & {$2(2\sqrt{2}-1)$} & 3.6569 & 3.6569 & 6 &   3.6569 &   2 &   2 &   2 &  3.6569 &    2 &    2 &    2\\
5 & 3 & 4.8885 & 4.8885 & 4.8885 & 7 &   4.6569 &   4.6569 &   4.6569 &   3.2097 & 4.6569 &    4.6569 &    4.6569 &    3.0187 \\
6 & 3 & 4.6569 & 4.6569 & 4.6617 & 7 &   4.6569 &   4.6569 &   3.0000 &   3.0000 & 4.6569  &  4.6569 &    3.0000 &    3.0000\\
7 & 4 & 20/3 & 6.6667 & 6.6667 & 10 &   5.6569 &   5.6569 &   5.6569 &   4.0000 & 5.6569 &    5.6569 &    5.6569 &    4.0000\\
8 & 4 & 6.6667 & 6.6667 & 6.6667 & 8 &   5.6569 &   5.6569 &   5.6569 &   4.0000 & 5.6569  &  5.6569 &    5.6569 &    4.0000 \\
9 & 4 & 5.6569 & 5.6569 & 5.6569 & 8 &   5.6569 &   4.0000 &   5.6569 &   4.0000 & 5.6569  &  4.0000 &    5.6569 &    4.0000 \\
10 & 4 & 4 & 4 & 5.3211 & 20/3 &   4 &   4 &   4 &   4 & 4 &    4 &    4 &    4\\
11 & 4 & 5.6569 & 5.6569 & 5.6569 & 8 &   4.0000 &   4.0000 &   5.6569 &   4.0000 & 4.0000  &  4.0000 &    5.6569 &    4.0000\\
12 & 4 & 5.6569 & 5.6569 & 5.6569 & 8 &   4.3695 &   4.3695 &   5.6569 &   4.2830 & 4.0085$^*$  &  4.0088$^*$ &    5.6569 &    4.0007$^*$\\
13 & 4 & 5.6569 & 5.6569 & 5.6569 & 8 &   5.6569 &   4.0000 &   5.6569 &   4.0000 & 5.6569  &  4.0000 &    5.6569 &    4.0000 \\
14 & 4 & 5.6569 & 5.6569 & 5.6569 & 8 &   4.0000 &   4.0000 &   5.6569 &   4.0000 &  4.0000  &  4.0000 &    5.6569 &    4.0000\\
15 & 4 & 6.0000 & 6.0000 & 6.0000 & 8 &   5.6569 &   4.4517 &   5.6569 &   4.2243 & 5.6569 &    4.0095$^*$ &    5.6569 &    4.0000\\
16 & 4 & 6.1289 & 6.1289  & 6.1289 & 8 &   5.6569 &   5.6569 &   5.6569 &   4.0000 & 5.6568  &  5.6569 &    5.6569&    4.0000\\
17 & 4 & 5.6569 & 5.6569 & 5.6569 & 8 &   4.0000 &   5.6569 &   5.6569 &   4.0000 & 4.0000 &   5.6569 &    5.6569&    4.0000\\
18 & 4 & 5.7538 & 5.7538 & 5.7538 & 8 &   5.6569 &   4.3130 &   4.3130 &   4.2247 & 5.6569  &  4.0000   & 4.0000&    4.0000\\
19 & 4 & 5.7829 & 5.7829 & 5.7829 & 8 &   5.6569 &   5.6569 &   4.3063 &   4.1865 & 5.6569  &  5.6569   & 4.0000&    4.0000 \\
20 & 4 & 6.4853 & 6.4853 & 6.4853  & 10 &  6.4853 &   4.5000 &   4.6903 &   4.1328  & 6.4853&    4.5000&    4.6847&    4.0000\\
21 & 4 & 5.9555 & 5.9555 & 5.9555 & 60/7 &   5.6569 &   5.6569 &   5.6569 &   4.1749 & 5.6569&    5.6569&    5.6569&    4.0000\\
22 & 4 & 6.1980 & 6.1980 & 6.1980 & 8 &   5.6569 &   5.6569 &   5.6569 &   4.2748 & 5.6569  &  5.6569 &    5.6569 &    4.0000\\
23 & 4 & 4.6847 & 4.7754 & 5.2939 & 8 &   4.5000 &   4.5000 &   4.6847 &   4.1135 & 4.5000  &  4.5000 &    4.6847 &    4.0000\\
24 & 5 & 7.9401 & 7.9401 & 7.9401 & 31/3 &   6.6569 &   6.6569 &   6.6569 &   5.2372 & 6.6569  &  6.6569 &    6.6569 &    5.0000\\
25 & 5 & 6.8243 & 6.8243 & 6.8415 & 31/3  &   6.6569 &   6.6569 &   6.4272 &   5.1652 & 6.6569  &  6.6569 &    6.4272 &    5.0000 \\
26 & 5 & 7.9282 & 7.9282 & 7.9282 & 31/3 &   6.4272 &   6.4272 &   6.4272 &   5.1819 & 6.4272  &  6.4272 &    6.4272 &    5.0000 \\
27 & 5 & 6.9547 & 6.9547 & 6.9588 & 31/3 &   6.4272 &   6.6569 &   6.6569 &   5.1808 & 6.4272 &    6.6569 &    6.6569 &    5.0000 \\
28 & 6 & 9.9098 & 9.9098 & 9.9098 & 14 &   9.3137 &   7.4272 &   7.4272 &   6.2123 &  9.3137  &  7.4272 &    7.4272 &    6.0000 \\
29 & 6 & 9.3137 & 9.3137 & 9.3137 & 14 &   9.3137 &   7.4272 &   7.4272 &   6.1624 & 9.3137  &  7.4272 &    7.4272 &    6.0000 \\
30 & 6 & 9.3137 & 9.3137 & 9.3137 & 14 &   9.3137 &   7.4272 &   7.4272 &   6.1723 & 9.3137  &  7.4272 &    7.4272 &    6.0000 \\
31 & 6 & 7.8043 & 7.8043 & 7.9226 & 12 &   7.6569 &   7.4272 &   7.4272 &   6.1866 & 7.6569 &   7.4272 &    7.4272 &    6.0000\\
32 & 6 & 8.1516 & 8.1516 & 8.1754 & 12 &   7.6569 &   7.6569 &   7.4272 &   6.2086 & 7.6569 &  7.6569 &    7.4272 &    6.0000\\
33 & 6 & 9.7899 & 9.7899 & 9.7899 & 12 &   7.6569 &   7.6569 &   7.6569 &   6.3217 & 7.6569 &  7.6569 &    7.6569 &    6.0000\\
34 & 6 & 8.2515 & 8.2515 & 8.2723 & 12 &   7.6569 &   7.4272 &   7.4272 &   6.2444 & 7.6569  &  7.4272 &    7.4272 &    6.0000\\
35 & 6 & 7.8553 & 7.8553 & 8.0776  & 12 &   7.4272 &   7.4272 &   7.4272 &   6.1794 & 7.4272 &   7.4272 &    7.4272 &    6.0000\\
36 & 6 & 9.4614 & 9.4614 & 9.4614 & 14 &   9.3137 &   7.4272 &   7.4272 &   6.1904 & 9.3137 &   7.4272 &    7.4272 &    6.0000 \\
37 & 6 & 9.3137 & 9.3137 & 9.3137 & 14 &   9.3137 &   7.4272 &   7.4272 &   6.1817 & 9.3137  &  7.4272 &    7.4272 &    6.0000 \\
38 & 6 & 9.3137 & 9.3137 & 9.3137 & 14 &   9.3137 &   7.4272 &   7.4272 &   6.1627 & 9.3137  &  7.4272 &    7.4272 &    6.0000\\
39 & 6 & 9.3253 & 9.3253 & 9.3253 & 12 &   7.6569 &   7.6569 &   7.6569 &   6.4378 & 7.6569 &  7.6569 &   7.6569 &   6.0000\\
40 & 6 & 8.1298 & 8.1298 & 8.1458 & 12 &   7.4272 &   7.6569 &   7.4272 &   6.2677 & 7.4272  &  7.6569 &    7.4272 &    6.0000\\
41 & 7 & 10.3677 & 10.3735 & 10.3769 & 15 &  10.3137 &  10.3137 &   8.4272 &   7.2012 & 10.3137 &  10.3137 &    8.4272 &    7.0000\\
42 & 8 & 13.0470 & 13.0470 & 13.0470  & 16 &  10.9852 &  10.9852 &  11.3137 &   8.2933 & 10.9852 &  10.9852 &   11.3137 &    8.0000\\
43 & 8 & 11.3137 & 11.3137 & 11.3137 & 16 &  10.9852 &   9.4272 &  11.3137 &   8.2481 & 10.9852  &  9.4272 &   11.3137 &    8.0000\\
44 & 8 & 12.9706 & 12.9706 & 12.9706 & 20 &  12.9706 &   9.3693 &   9.3693 &   8.2812 & 12.9706 &    9.3693 &    9.3693 &    8.0000 \\
45 & 8 & 12.9706 & 12.9706 & 12.9706 & 20 &  12.9706 &   9.3693 &   9.3693 &   8.2675 & 12.9706  &  9.3693 &    9.3693 &    8.0000 \\
46 & 10 & 12.9852 & 12.9852 &  13.2668 & 62/3 &  12.8543 &  12.8543 &  12.9852 &  10.4006 & 12.8543 &   12.8543 &   12.9852  & 10.0000\\
\hline
\end{tabular}
\caption{Maximal value (violation) of the Bell polynomials (inequalities) presented in Ref.~\cite{sliwa} for various sets of correlations. From the second column to the sixth column, we have, respectively, the maximal value attainable by correlations from the local set $\L$, the quantum set $\Q$, the almost quantum set $\AQ$, the 2nd level of the NPA relaxations $\Q_2$ and the  no-signalling set $\NS$. The next four columns show the maximal value attainable by {subsets of} almost quantum correlations that arise from a moment matrix {having} positive partial transposition~\cite{Peres:PPT} (PPT) with respect to certain bipartition (for example, $\AQ^{\tiny T_A}_1$ means that the corresponding moment matrix $\chi_1$ is further subjected to the PPT constraint $\chi_1^{\tiny T_A}\succeq 0$, while $\AQ_1^{\tiny T_{all}}$ means that the PPT condition is imposed for all bipartitions concurrently). The {last} four columns show the maximal value attainable by correlations described by a higher-level refinement of $\AQ$ (specifically, local level six as described in Ref.~\cite{Moroder:PRL:2013}) when subjected to further PPT constraints as described above (see text in Sec.~\ref{Sec:Refine} for more details). We see that two out of the 46 inequalities (n$^o$ 23 and 41) display a gap between the maximal violation from $\Q$ and $\AQ$, hence witnessing the post-quantumness of the latter set. Note that from the relations among the violations we can recognise that Mermin inequality~\cite{mermin} belongs to family n$^o$ 2, the party-lifting~\cite{lifting} of the CHSH inequality {---whose maximal quantum violation inherits directly from that of the CHSH inequality---} belongs to family n$^o$ 4 , and the GYNI~\cite{gyni} belongs to n$^o$ 10. Entries marked with $^*$ represent bounds which are not known to be saturated by any three-qubit states featuring the same kind of positive partial transposition.} \label{Bounds322}
\end{table*}

\section{Almost quantum and the NPA hierarchy}
\label{Sec:NPA}

Characterisation of the set of quantum correlations $\Q$ is a notoriously difficult problem. A first systematic approach to this problem was introduced by {NPA} \cite{npa} where they provided an algorithmic characterisation of $\Q$ via a hierarchy $\Q_1 \supseteq \Q_2 \supseteq \ldots \supset \Q_k$ of (postquantum) subsets to $\NS$ which converges to $\Q$ in the asymptotic limit of $k \rightarrow \infty$ (see also Ref.~\cite{Doherty:qmp}). Algorithmically, each set $\Q_k$ (for positive integer $k$) is characterised by the set of feasible solutions to an SDP (see Appendix~\ref{App:NPA} for details), which demands that a {Hermitian} matrix $\Gamma_k$---whose entries being {\em all} the {moments} (expectation values) up to order $2k$---can have only non-negative eigenvalues. Importantly, some of these moments {correspond precisely to} those joint conditional probabilities $p(abc|xyz)$ between measurement outcomes that can be estimated directly from a Bell-type experiment. For instance, in an $n$-partite Bell scenario, each joint conditional probability given in $\vecP$ is a moment of order $n$. If $\vecP\in\Q$, then as mentioned in Sec.~\ref{Sec:Local}, $\vecP$ {[as well as other (higher-order) moments]} can be accountable for through Born's rule, thereby resulting in a moment matrix that has only non-negative eigenvalues~\cite{npa} {(see also Appendix~\ref{App:NPA})}. 

{In general, however, the non-negativity of a moment matrix $\Gamma_k$ is not sufficient to guarantee that $\vecP$ has a quantum origin. In other words, for {\em any} given $k$, all given $\vecP$ where the corresponding $\Gamma_k$ (which contains $\vecP$ as entries) can be made non-negative form a superset of $\Q$, which is precisely the NPA set $\Q_k$.}
In particular, since the moment matrix $\Gamma_{k+1}$ associated with the characterisation of $\Q_{k+1}$ contains the moment matrix $\Gamma_k$ associated with the characterisation of $\Q_{k}$ as a submatrix, it thus follows that $\Q_{k+1}\subseteq\Q_k$ for all positive integer $k$.

Instead of the definition given in Sec.~\ref{Sec:Local} via permutations, the almost quantum set $\AQ$ can equivalently be defined~\cite{aq} as the set of $\vecP$ that can be characterised using an analogous SDP involving a moment matrix of appropriate size. In fact, it turns out~\cite{miguel} that this SDP characterization of $\AQ$ is precisely the {\em lowest level} superset characterisation of $\Q$ introduced by Moroder {\em et al.}~\cite{Moroder:PRL:2013}. In the bipartite case, it is thus known~\cite{npa,aq} that\footnote{The set $\Q_{1+AB}$ is an intermediate level in the NPA hierarchy, whose moment matrix $\Gamma_{1+AB}$ {has rows and columns  labelled} by words which involve single party operators as well as all the two party ones (see Appendix \ref{App:AQ}).} $\Q_1 \supset \AQ=\Q_{1+AB} \supset \Q_2$, and more generally in the $n$-partite case, we must have $\AQ\supseteq \Q_n$. What about the relationship between $\Q_{n-1}$ and $\AQ$? Does the strict inclusion that holds in the bipartite ($n=2$) case also generalise to the situation when $n>2$?

To answer this question, let us now focus our attention on the fourth and fifth columns of Table~\ref{Bounds322}, which shows the maximum violation of {the 46 facet} Bell inequalities presented in Ref.~\cite{sliwa}, respectively, for $\vecP\in\AQ$ and $\vecP\in\Q_2$.
We see that for some of these non-trivial tripartite Bell inequalities, there exists correlations $\vecP$ in $\Q_2$ that are more nonlocal (in the sense of giving a stronger Bell violation) than those restricted to the almost quantum set. From here, we can thus conclude that $\Q_2\not\subseteq\AQ$ in this tripartite Bell scenario. Could it then be that $\AQ\subset\Q_2$? Inspired by the construction given in Ref.~\cite{aq}, we found that the following tripartite Bell inequality answers this in the negative:
\begin{widetext}
\begin{equation*}
 \sum_{i\in{A,B,C}}  \left[\tfrac{30}{31}p_i (0|0) -\tfrac{167}{9} p_i(0|1)\right] - \sum_{ij\in\{AB,AC,BC\}} \tfrac{74}{11} p_{ij}(00|00)  + \tfrac{174}{11} \left[p_{AB}(00|01) + p_{AB} (00|10)\right] + \tfrac{244}{23} p_{AB}(00|11) \stackrel{\L}{\leq} \tfrac{30}{31}\,.
\end{equation*} 
\end{widetext}
Specifically, the set of tripartite almost quantum correlations $\AQ$ achieve a maximal violation of $1.0232$ for this Bell inequality, whereas correlations in $\Q_2$ only yields a maximum violation of $0.9724$. This implies that there exist tripartite almost quantum correlations that do not belong to $\Q_2$. 

Our results then show that in this tripartite scenario, neither the almost-quantum set {contains} the second level of the NPA set nor the other way around, i.e., we have both $\AQ \not\subseteq \Q_2$ and $\AQ \not\supseteq \Q_2$. In other words, for general $n\ge 3$, the inclusion relation $\Q_{n-1}\supseteq\AQ$ need not hold true.

\section{Refinements to the almost quantum set}
\label{Sec:Refine}

For any correlation that arises from local measurements on a  quantum state $\rho$, it was noted~\cite{Moroder:PRL:2013} that when the expectation values $\{\expect{O_i^\dag O_j}_\rho\}$ are organized as entries of an NPA moment matrix, i.e., 
\begin{equation}\label{Eq:Gamma_k}
	\Gamma_k=\sum_{i,j} \ket{i}\expect{O_i^\dag O_j}_\rho \bra{j} 
\end{equation}
the corresponding $\Gamma_k$ (for any $k$) can be seen as the result of a global, completely positive map {$\Lambda$} acting on $\rho$, that is, $\Gamma_k=\Lambda(\rho)$. 
Moreover, for $k=n\ell$ with $n$ being the number of parties and $\ell$ being a positive integer, a particular submatrix of $\Gamma_k$, which we denote by $\chi_\ell$ (see Appendix~\ref{App:AQ}), can even be seen as the result of a separable, completely positive map acting on the underlying state $\rho$, i.e.,  $\chi_\ell=\Lambda_1\otimes\Lambda_2\otimes\cdots\Lambda_n(\rho)$. See Ref.~\cite{Moroder:PRL:2013} for details.

For reasons that will become obvious below, let us denote by $\AQ_\ell$ the set of $\vecP$ compatible with the constraint of $\chi_\ell\succeq 0$, the positivity of the NPA moment {\em submatrix} as mentioned above {(see Appendix~\ref{App:AQ} for details)}. As with the original NPA hierarchy of moment matrices, the series of structured moment matrices {$\chi_1,\chi_2,\ldots,\chi_\ell$ also give rise to a hierarchy of superset $\AQ_1,\AQ_2,\ldots,\AQ_\ell$ characterization} of $\Q$ which asymptotically converges~\cite{Moroder:PRL:2013} to $\Q$ as $\ell\to\infty$. Similarly, let us denote by $\AQ_\ell\tp$ the set of correlations compatible with both the constraints of $\chi_\ell\succeq 0$ and $\chi_\ell^\text{\tiny $T_i$}\succeq0$, where {$(.)^{\tiny T_i}$} denotes partial transposition~\cite{Peres:PPT} with respect to party $i$. As was shown in Ref.~\cite{Moroder:PRL:2013}, as $\ell\to\infty$, the series of $\AQ_\ell\tp$ converge to $\Q\tp\subset\Q$, i.e., the set of quantum correlations arising from a density matrix $\rho$ that has the property of $\rho\tp\succeq 0$.

Incidentally, $\AQ_1$ as explained above is exactly~\cite{miguel} the equivalent formulation of the almost quantum set that we have alluded to in Sec.~\ref{Sec:NPA} (see also Ref.~\cite{aq}), i.e., $\AQ=\AQ_1$. All the higher level relaxations of $\Q$ discussed in Ref.~\cite{Moroder:PRL:2013} can thus be seen as some kind of refinements to the almost quantum set. Let us denote this by {$\P=\{A,B,C\}$.} Then, one can similarly see $\AQ_1\tp$ for different bipartitions of {$\P$} into disjoint subsets $\mathcal{A}$ and {$\P\setminus\mathcal{A}$} as different refinements of the almost quantum set which allows only a ``PPT" resource between the group of parties in $\mathcal{A}$ and those in ${\P}\setminus\mathcal{A}$. {In the rest of this paper, we thus posit that $\AQ_1\tp$ is the subset of $\AQ$ where only a separable resource with respect to the bipartition of party $i$ vs $\P\setminus\{i\}$ is available.}

In Table~\ref{Bounds322}, we  have included in the seventh to the ninth columns the maximal value of all the inequalities given in Ref.~\cite{sliwa} when only such weaker resources from the almost quantum {set, i.e., $\vecP\in\AQ_1\tp$ are allowed}. {To facilitate comparison, we have included in columns 11--13 of Table~\ref{Bounds322}, respectively, the maximal value of each inequality attainable by $\vecP\in\AQ_6\tp$ for $i\in\{A,B,C\}$.\footnote{Note that at this level of the hierarchy, one already needs to deal with a moment matrix of size $2197\times 2197$, which involves 8192 distinct moments.}  Evidently, these values illustrate how these higher-level refinements of the almost quantum set differ from $\AQ$  when only these weaker forms of ``biseparable" resources are allowed.  It is worth noting that apart from a few exceptional cases (which we marked with an asterisk $^*$ in the table), all these bounds are saturated (within the numerical precision of the solver) by some biseparable three-qubit states\footnote{{The recent work of Ref.~\cite{Adan16} independently came to this conclusion that the maximal quantum violation of many of these Bell inequalities can already be attained using a biseparable three-qubit state.}}. }

Interestingly, among those nontrivial inequalities, the maximal value attainable by the almost quantum set $\AQ$ {(column four, Table~\ref{Bounds322})} for 19 of these inequalities\footnote{These concern inequalities 3, 4, 6, 9-14, 17, 20, 29, 30, 37, 38, 43-46.} can already be attained (within {the numerical precision of  the solver}) by $\vecP\in\AQ_1\tp$ for some {$i\in\P$ (columns seven to nine, Table~\ref{Bounds322}). Moreover,} the same applies to the maximal quantum value {(column three, Table~\ref{Bounds322})}  of these inequalities and what can be achieved assuming only quantum resources that are {biseparable (columns eleven to thirteen, Table~\ref{Bounds322})} with respect to the same bipartition. {Intriguingly,} for inequality 23, the maximal almost-quantum value when assuming such weaker resource across the bipartition of $A|BC$ {is even} closer to the maximal quantum value than to the maximal almost-quantum value. 

{Apart from fundamental interests, the fact that the bounds presented in the last four columns can be (mostly) saturated by biseparable quantum states also has pragmatic implications. In particular, for} all those Bell inequalities where there is a gap between its maximal quantum violation (column three) and the maximum value of $\AQ_6\tp$ (columns eleven to thirteen), we can possibly use the extent to which the inequality is violated to conclude directly from the measurement statistics (i.e., in a device-independent~\cite{RMP} manner) the presence of a genuine {tripartite} entangled state. For example, with the appropriate bound in place, the quantum violation of inequality 7~\cite{DIWED} can be used as a device-independent witness for genuine tripartite entanglement~\cite{DIEW}:
\begin{equation}
\sum_{x,y,z=0}^1E(xyz)-4E(111) \stackrel{\text{\tiny 2-prod.}}{\leq} 4\sqrt{2}\,,
\end{equation}
where the superscript ``2-prod." here means that the bound holds for {\em any} 2-producible~\cite{k-prod} quantum states (or equivalently, tripartite but biseparable quantum states). Note also that in some cases, such as for inequality 20, quantum violation of the distinct biseparable bounds (for the different bipartitions) can even be used to certify the presence of entanglement across {\em particular} bipartition(s).

Finally, let us focus on the tenth column and the fourteenth column of Table~\ref{Bounds322} which list, respectively, the maximal value of $\AQ_1$ and $\AQ_6$ when we assume the weakest form of {``biseparable"} resource, namely, one that has a moment matrix that is {concurrently PPT across all three bipartitions}. Clearly, in the case of $\AQ_6$, we see that apart from inequality 5 and inequality 12, no quantum violation is possible when only these weakest forms of resources are available.\footnote{Note that $\AQ_6\supseteq\Q$. Thus if an inequality cannot be violated by (fully) biseparable higher-level almost-quantum correlation, it also cannot be violated by (fully) biseparable quantum correlation.} In stark contrast, the almost quantum correlation can still violate 32 out of the 45 nontrivial Bell inequalities even if only such weakest form of resources are available. In the quantum setting, the violation of a Bell inequality due to such resources has been used to construct a counterexample~\cite{Vertesi} to the tripartite Peres conjecture~\cite{PeresConj}. If one can give an analogous interpretation of such a violation in the almost-quantum setting, then it would suggest that an undistillable but nonlocal almost-quantum resource in the tripartite setting is considerably more common than it is in the quantum setting.

\section{Conclusions}
\label{Sec:Conclusion}

In this work we explored properties of the set of almost quantum correlations and their refinements in the simplest tripartite Bell scenario, namely one with two dichotomic measurements per party. First we computed the maximum violations of all the facet Bell inequalities that define the set of classical correlations in this scenario, respectively, by the set of almost quantum correlations and by the set of quantum correlations themselves. We found that the difference between these two sets may indeed be witnessed by their violation of these Bell inequalities. This is in strong contrast with its bipartite counterpart, where all values of the CHSH Bell inequality achievable by almost quantum correlations have a corresponding quantum realisation. Our findings highlight how complex and rich multipartite Bell scenarios are, even in their simplest form. 

Then, we asked how the set of almost quantum correlations compares to other sets that include postquantum correlations. In particular, we focused on the NPA hierarchy for {this tripartite scenario}, where it is already known that $\Q_1 \supseteq \AQ \supseteq \Q_3$. In this work we proved that there is no such inclusion relation between the second level of NPA ($\Q_2$) and the almost quantum correlations, that is, we showed that neither  $\AQ \subseteq \Q_2$ nor $\AQ \supseteq \Q_2$. More generally, this implies that in an $n$-partite scenario where $n>2$,  neither $\AQ \subseteq \Q_{n-1}$ nor $\AQ \supseteq \Q_{n-1}$.

Finally, we investigated refinements of the almost quantum sets as characterised by the hierarchies of semidefinite programs proposed in Ref.~\cite{Moroder:PRL:2013}. In particular, by positing that a ``biseparable" almost-quantum resource is captured through the additional requirement of positive partial transposition~\cite{Peres:PPT} on the corresponding moment matrices, we could pinpoint some subtle similarities and differences between the quantum set and the almost quantum set of correlations.  For example, a significant fraction of these facet Bell inequalities can already be violated maximally for both sets of correlations using a biseparable resource. However, if we demand that the resource is simultaneously biseparable with respect to all bipartitions, then quantum correlations apparently can no longer violate most (but 2) of these inequalities, while the almost quantum correlations can still often provide an advantage over classical resources. Of course, in the light of the frameworks discussed in Refs.~\cite{DIEW,DIWED,Flo}, the bounds that we have computed are also useful as device-independent~\cite{RMP} witnesses for nontrivial properties of the underlying quantum systems.

Our results also open the door to several open questions. {For example, it} would be interesting to find an interpretation as a game or a physical principle of those tight Bell inequalities that witness almost quantum correlations beyond quantum theory. We believe this will shed light on the problem of characterising quantum correlations from basic principles.  Finally, it would be interesting to explore protocols to check violations of IC via these postquantum almost quantum correlations. 

\begin{acknowledgments}
We thank Jean-Daniel Bancal, Ad\'an Cabello, Matty Hoban, Miguel Navascu\'es, Paul Skrzypczyk, Sandu Popescu, {Elie Wolfe for fruitful discussions, and  Denis Rosset for spotting a typo in en earlier version of Table~\ref{Bounds322}.} We are particularly grateful to Zhen-Peng Xu for bringing to our attention issues in an earlier version of this manuscript. {ABS thanks Jean-Daniel Bancal for sharing his codes that facilitate the computation of some of the SDP bounds presented and} acknowledges financial support from ERC AdG NLST and Perimeter Institute for Theoretical Physics. Research at Perimeter Institute is supported by the Government of Canada through the Department of Innovation, Science and Economic Development Canada and by the Province of Ontario through the Ministry of Research, Innovation and Science. YCL acknowledges financial support by the Ministry of Education, Taiwan, R.O.C., through ``Aiming for the Top University Project" granted to the National Cheng Kung University (NCKU), and the Ministry of Science and Technology, Taiwan (Grant No.104-2112-M-006-021-MY3).

\end{acknowledgments}

\appendix

\section{Almost quantum and the NPA hierarchy as semidefinite programs}
\label{App}

In this {Appendix} we present the definition of the NPA hierarchy as outer approximations to the quantum set of correlations, as well as the SDP formulation of the almost quantum set. For the sake of clarity we focus on the Bell scenario explored in this work, i.e., the tripartite scenario involving two dichotomic measurements per party. For a general and the precise formulation of NPA as well as the almost quantum set, we refer the reader to Ref.~\cite{npa} and Ref.~\cite{aq} respectively. 

\subsection{The NPA hierarchy}
\label{App:NPA}

Let us start with the hierarchy of the {superset approximations to $\Q$} due to NPA. 
Each level of the NPA set $\Q_k$ is associated with a moment matrix $\Gamma_k$ with rows and columns labelled by specific sets of `words' (operators). For the $k$-th level, each label involves {the product} of {\em at most} $k$ projectors (which may all belong to the same party). For instance, in the tripartite scenario studied here the labels for $\Q_1$ are\footnote{In general, it was shown in Ref.~\cite{npa} that it is sufficient to consider the projectors corresponding to all but the last outcome. {In the following discussion, we shall thus consider only projectors associated with the 0-th outcome of each party.}}
\begin{align*}
S_1 = \openone \cup \{\Pi_{0|x},\Pi_{0|y},\Pi_{0|z}\}_{x,y,z},
\end{align*}
and the labels for $\Q_2$ are
\begin{align*}
S_2 =& S_1 \cup \{\Pi_{0|x}\Pi_{0|y},\Pi_{0|y}\Pi_{0|z},\Pi_{0|x}\Pi_{0|z}\}_{x,y,z}, \\
& {\cup \{\Pi_{0|x}\Pi_{0|x'},\Pi_{0|y}\Pi_{0|y'},\Pi_{0|z}\Pi_{0|z'}\}_{\substack{x,y,z,x',y',z'}},}
\end{align*}
{where $x\neq x'$, $y\neq y'$, $z\neq z'$.}

For any given state $\rho$ and operators $O_i, O_j\in S_k$, the expectation value $\expect{O_i^\dag O_j}=\text{tr}(\rho\,O_i^\dag O_j)$ is well-defined. In particular, if $\{\ket{i}\}$ is a set of orthonormal basis vectors, then
\begin{equation}\label{Eq:Gamma_k2}
	\Gamma_k=\sum_{i,j} \ket{i} \expect{O_i^\dag O_j} \bra{j} 
\end{equation}
is a matrix of expectation values, dubbed an NPA {\em moment matrix} of level $k$. A generic moment matrix satisfies a number of properties. Firstly, 
for any vector $\ket{v}$, it follows from Eq.~\eqref{Eq:Gamma_k2} that  $\bra{v}\Gamma_k\ket{v}=\sum_{i,j} \braket{v}{i}\braket{j}{v} \expect{O_i^\dag O_j}= \tr{(\sum_{i} v_iO_i)^\dag (\sum_{j}v_jO_j)} \ge 0$. Thus, a {legitimate quantum} moment matrix must be positive semidefinite, or in other words, having only non-negative eigenvalues:
\begin{equation}\label{Eq:PSD}
	\Gamma_k \succeq 0.
\end{equation}
 
In evaluating the entries of a moment matrix, one assumes, as with quantum theory, that operators belonging to different parties commute, i.e., $[\Pi_{a|x},\Pi_{b|y}]=[\Pi_{a|x},\Pi_{c|z}]=[\Pi_{b|y},\Pi_{c|z}]=0$. Thus, certain entries of $\Gamma_k$ arising from different combinations of $O_i$, $O_j$, $O_i'$, and $O_j'$ may be identical.  As an example, consider $k=2$ and $O_i=\Pi_{a|x}\Pi_{b|y}$, $O_j=\Pi_{a|x}\Pi_{c|z}$, we thus see that $O_i^\dag O_j = (\Pi_{a|x}\Pi_{b|y})^\dag \Pi_{a|x}\Pi_{c|z} =\Pi_{a|x}^2\Pi_{b|y}\Pi_{c|z} = \Pi_{a|x}\Pi_{b|y}\Pi_{c|z} =O_i'^\dag O_j'$ {if} $O_i'=\Pi_{a|x}$, $O_j'=\Pi_{b|y}\Pi_{c|z}$ and the second last equality follows from the projective nature of $\Pi_{a|x}$. Denoting by $\Gamma_k(i,j)$ the $(i,j)$-th entry of $\Gamma_k$, then we shall collectively refer to such equality constraints as:
\begin{equation}\label{Eq:EqualityConstraint}
	\Gamma_k(i,j)=\Gamma_k(i',j')\quad\forall\,\,O_i^\dag O_j=O_i'^\dag O_j'.
\end{equation}

Lastly, note that if $O_i$ and $O_j$ are such that $O_i^\dagger O_j$ involves at most one projector per party, then the corresponding expectation value corresponds to joint distribution that can be estimated from an experiment. In particular, if the projector of a party does not appear in the product $O_i^\dag O_j$, then the resulting expectation value is simply a marginal distribution that does not involve the corresponding party. Explicitly, this means that the entries of the moment matrix satisfy:
\begin{equation}\label{Eq:correlation}
\begin{split}
	\expect{\openone^\dag\openone} &= 1\,,\\ 
	\expect{\openone^\dag\Pi_{a|x}} &= p_A(a|x)\,,\\
	\expect{\Pi_{a|x}^\dag\Pi_{b|y}} &= p_{AB}(ab|xy) \,,\\
	\expect{\Pi_{a|x}^\dag\Pi_{b|y}\Pi_{c|z}} &= p(abc|xyz) \,,
\end{split}
\end{equation}
and similar relations for the other parties' marginals, where $p_A$ and $p_{AB}$ are the one and two party marginals of $p(abc|xyz)$. Notice that for $k\ge 2$, one can always find $O_i,O_j\in S_k$ such that the above conditional distributions are recovered as entries of $\Gamma_k$.

Conversely, whenever a matrix of the form of Eq.~\eqref{Eq:Gamma_k} satisfies the constraints given in Eq.~\eqref{Eq:PSD} and Eq.~\eqref{Eq:EqualityConstraint}, it is a legitimate moment matrix of level $k$. The set of all $\vecP=\{p(abc|xyz)\}$ compatible with these constraints therefore define the $k$-th level of the (postquantum) NPA set, $\Q_k$. Determining if a given $\vecP$ belongs to $\Q_k$ therefore amounts to finding a positive semidefinite matrix (the moment matrix $\Gamma_k$) that satisfies the linear equality constraints of Eq.~\eqref{Eq:EqualityConstraint} and Eq.~\eqref{Eq:correlation}, which is therefore {an SDP}. Likewise, finding the maximal value of a (linear) Bell inequality for correlations $\vecP\in\Q_k$ for any integer $k\ge 2$ amounts to maximizing a certain linear combination of entries in a positive semidefinite matrix that is subjected to the linear equality constraints of Eq.~\eqref{Eq:EqualityConstraint} and Eq.~\eqref{Eq:correlation}, which is also an SDP.

\subsection{The almost quantum set and its refinements}
\label{App:AQ}

We now switch to the almost quantum set $\AQ$ and its refinements. Instead of the definition given in Sec.~\ref{Sec:Local} via permutations, {the almost} quantum set can be equivalently formulated~\cite{aq,miguel} as correlations $\vecP$ in an intermediate NPA set. In particular, for the tripartite Bell scenario that we consider in this paper, the almost quantum set of correlations $\AQ$ is precisely the set of $\vecP$ compatible with a moment matrix labelled by  `words' (operators) {defined in} the following set
\begin{equation}\label{Eq:Saq}
\begin{split}
	S_{aq} =& \openone \cup \{\Pi_{0|x},\Pi_{0|y},\Pi_{0|z}\}_{x,y,z} \\
	& \cup \{\Pi_{0|x}\Pi_{0|y},\Pi_{0|y}\Pi_{0|z},\Pi_{0|x}\Pi_{0|z}\}_{x,y,z}, \\
	& \cup \{\Pi_{0|x}\Pi_{0|y}\Pi_{0|z}\}_{x,y,z}.
\end{split}
\end{equation}
The resulting moment matrix, which we denote by $\chi_1$ {thus has exactly the same form as the right-hand-side of Eq.~\eqref{Eq:Gamma_k2}, and is therefore a submatrix of $\Gamma_3$ of size $27\times 27$.}

As with an ordinary NPA set $\Q_k$, all $\vecP$ such that the corresponding $\chi_1$ is compatible with the {analogous} constraints of Eq.~\eqref{Eq:PSD}, Eq.~\eqref{Eq:EqualityConstraint} and Eq.~\eqref{Eq:correlation} is a convex set and, thanks to the equivalent formulation~\cite{aq,miguel}, gives rise to precisely the almost quantum set $\AQ$. Thus, as with the membership test of $\Q_k$, determining if a given $\vecP$ is a member of $\AQ$ amounts to finding a positive semidefinite {(moment)} matrix that satisfies all the aforementioned linear equality constraints, and therefore an SDP.

To consider refinements of the almost quantum set defined in Ref.~\cite{Moroder:PRL:2013}, let us note that the list of operators given in Eq.~\eqref{Eq:Saq} can be alternatively written as:
\begin{equation}\label{Eq:Saq-equiv}
\begin{split}
	S_{aq} =& \{\openone, \Pi_{0|x}\}_x \times \{\openone, \Pi_{0|y}\}_y \times \{\openone, \Pi_{0|z}\}_z
\end{split}
\end{equation}
where $\times$ here represents the Cartesian product. We can thus see that the `words' defining the almost quantum set are those formed by the Cartesian product of local operators that are up to order 1. The next level refinement of this set then involves {the} Cartesian product of local operators that are up to order 2, as specified by the following set of words:
\begin{equation}\label{Eq:Saq2}
\begin{split}
	S_{aq,2} = &\{\openone, \Pi_{0|x}, \Pi_{0|x}\Pi_{0|x'}\}_{x,x'}\\
	 &\times \{\openone, \Pi_{0|y}, \Pi_{0|y}\Pi_{0|y'}\}_{y,y'} \\
	&\times \{\openone, \Pi_{0|z}, \Pi_{0|z}\Pi_{0|z'}\}_{z,z'}.
\end{split}
\end{equation}
Even higher level refinements of the almost quantum set can be defined analogously. In general, the (local) level $\ell$ set of words are defined by the Cartesian product of local operators that are up to order $\ell$. We shall write the corresponding moment matrix as $\chi_\ell$ and the corresponding set of correlations $\vecP$ that are compatible with the {analogous} constraint of Eq.~\eqref{Eq:EqualityConstraint}, Eq.~\eqref{Eq:correlation} and $\chi_\ell\succeq0$ as $\AQ_\ell$. Thus, the membership test of any given $\vecP$ in relation to $\AQ_\ell$ is {an SDP, so is the computation of the maximal value of a (linear) Bell inequality for $\vecP\in\AQ_\ell$. In fact, it is straightforward to see that the same applies to those computations pertaining to the refined subsets $\AQ_\ell\tp$ of $\AQ_\ell$.}

\end{document}